\documentclass[aps,prl,reprint,longbibliography]{revtex4-1}

\pdfoutput=1

\usepackage{braket}
\usepackage{amssymb}
\usepackage{amsmath}
\usepackage{amsthm}
\usepackage{tikz}
\usetikzlibrary{arrows, calc}
\usepackage{xcolor}
\usepackage{graphicx}
\usepackage{makecell}
\usepackage{adjustbox}
\usepackage{appendix}
\usepackage{tablefootnote}
\usepackage{tabularx}
\usepackage{enumerate}

\hypersetup{colorlinks=true,linkcolor=color1,urlcolor=color2,citecolor=color3,anchorcolor=green}

 \usetikzlibrary{math}

\definecolor{color1}{RGB}{230,57,70}
\definecolor{color2}{RGB}{29,53,87}
\definecolor{color3}{RGB}{69,123,157}

\definecolor{cbRed}{RGB}{245, 121, 58}
\definecolor{cbPurple}{RGB}{169, 90, 161}
\definecolor{cbBlue}{RGB}{133, 192, 249}

\newcommand{\outline}[1]{}

\begin{document}

\title{A Compact Fermion To Qubit Mapping}
\author{Charles Derby}
\affiliation{Phasecraft Ltd.}
\affiliation{Department of Computer Science, University College London}
\author{Joel Klassen}
\email{joel@phasecraft.io}
\affiliation{Phasecraft Ltd.}


\begin{abstract}
Mappings between fermions and qubits are valuable constructions in physics. To date only a handful exist. In addition to revealing dualities between fermionic and spin systems, such mappings are indispensable in any quantum simulation of fermionic physics on quantum computers. The number of qubits required per fermionic mode, and the locality of mapped fermionic operators strongly impact the cost of such simulations. We present a novel fermion to qubit mapping which outperforms all previous local mappings in both the qubit to mode ratio, and the locality of mapped operators.






\end{abstract}

\maketitle

One of the most striking features of fermions is the non-locality of their state space. This non-locality is necessitated by their anti-symmetric exchange statistics --- the phase  of the wavefunction yielded by a fermion tracing a path past an even number of its counterparts differs from that yielded by a path past an odd number. However causality is preserved by parity superselection \cite{wick52}, which forbids superpositions of even and odd fermion number, preventing the direct measurement of these phase differences. 

A consequence of this non-locality is that any representation of fermionic systems on collections of local quantum systems, such as qubits or distinguishable spins, must introduce non-local structure\cite{LevinWen2003}. This is most readily seen in the Jordan-Wigner (JW) transform\cite{JordanWigner}, which maps fermionic creation ($a_i^{\dagger}$) and annihilation ($a_i$) operators, which create and annihilate a fermion at mode $i$ and satisfy the canonical anti-commutation relations 
\begin{align}
\{a_i^{\dagger}, a_j\}=\delta_{ij} \; , \; \{a_i^\dagger, a_j^\dagger \} = 0 \; , \;  \{a_i, a_j \} = 0 ,
\end{align}
to string-like Pauli operators
\begin{equation}
a_i^{\dagger} \rightarrow \frac{1}{2}Z_1 ... Z_{i-1} \left(X_i-i Y_i \right).
\end{equation}
Under this mapping even local observables conserving fermion parity, such as lattice hopping terms ($a^{\dagger}_i a_j + a^{\dagger}_ja_i$), are mapped to strings of Pauli operators which may be as large as the size of the system.

The JW transform is an example of a mapping between fermions and qubits. Such mappings describe a correspondence between states of fermions and states of qubits, or, equivalently, between fermionic operators and  multi-qubit operators. They are restricted to fermionic systems with a discrete set of modes, since qubits posses finite dimensional Hilbert spaces, and are typically applied to fermionic lattice models. Many mappings are tailored to specific lattices.

A potential application where fermion to qubit mappings would be indispensable is the simulation of fermions by quantum computers. The accurate simulation of fermions has long posed a fundamental challenge to classical computers. Since the conception of quantum computers it has been understood that one of their primary applications would be in addressing this challenge, with  substantial potential impact on a broad range of scientific disciplines. 

Using a fermion to qubit mapping, a fermionic Hamiltonian may be mapped to a qubit Hamiltonian $H=\sum_i H_i$, with the terms $H_i$ constituting tensor products of Pauli operators. A quantum computer can perform an effective simulation of the fermionic Hamiltonian by simulating time dynamics under $H$. 
 The primary strategy to do this is via a Trotter expansion, which consists of dividing the time evolution unitary into a product of short evolutions generated by the terms $H_i$ \cite{lloyd96, lloyd97, childs19}. This is followed by a further decomposition of these short evolutions into sequences of quantum gates. However the greater the number of qubits on which these individual Hamiltonian terms $H_i$ act --- ie the Pauli weight --- the more costly the circuit decomposition \cite{Havlicek2017,epsilonCircuts}.  Similar considerations also inform the performance of other quantum algorithms, such as VQE \cite{cade2019strategies}. Thus there has emerged a practical need to design fermion to qubit mappings which minimize the Pauli weight of commonplace fermionic operators. In particular those fermionic interactions which couple nearby fermionic modes --- ie geometrically local operators --- which feature prominently in physically realistic systems.

The JW transform performs poorly in this respect because all of the requisite fermionic non-locality is manifest in the observables, as opposed to the states --- the fermionic Fock states map directly to seperable binary  states. One may instead design a mapping which encodes the non-locality in the states, by mapping fermionic states into a highly entangled subspace of the multi-qubit system. In this way one can retrieve low weight qubit representations of geometrically local fermionic operators. We refer to such mappings as {\em local} mappings. 

There currently exist a handful of local mappings \cite{verstraete2005mapping,whitfield2016local, steudtner2019square,bravyi2002fermionic, setia2019superfast,jiang2018majorana}. A comparison of these mappings is given in Table~\ref{table:mappings}. Two terms which are ubiquitous in fermionic Hamiltonians are lattice hopping, and Coulomb interactions. The minimum upper bounds on the Pauli weights of these terms under any of these local mappings is 4 and 2 respectively. Furthermore, all of these local mappings employ approximately 2 or more qubits per fermionic mode.

In this letter we present a new local mapping that, when applied to square lattices, not only outperforms all existing local mappings in terms of Pauli weight, yielding for instance max weight 3 hopping terms, but also employs fewer than 1.5 qubits per mode. We expect these features to find significant use in near term quantum computing applications, where resources are limited. The mapping can be thought of as a modified toric code\cite{kitaev97, savary16} that condenses local pairs of particle excitations, yielding a low energy subspace which corresponds to a fermionic Hilbert space. For clarity we focus in this work on the square lattice, however the design scheme of the mapping may also be applied to other interaction graphs, yielding similar cost benefits. 


All local mappings encode fermionic states in a subspace of a multi-qubit Hilbert space via the formalism of stabilizer codes, by defining a set of mutually commuting Pauli operators (stabilizers) for which the subspace constitutes a common +1 eigenspace. There are two design strategies that all existing mappings employ. 
One strategy, employed by mappings presented in refs.~\cite{verstraete2005mapping,whitfield2016local, steudtner2019square}, leverages the Jordan-Wigner transform, and defines stabilizers that ``cancel out'' sections of the long strings, discarding a portion of the fermionic modes in the process. 

\outline{-what is a fermion to qubit mapping?}

The second strategy, employed by the mappings presented in refs.~\cite{bravyi2002fermionic, setia2019superfast,jiang2018majorana}, as well as the mapping presented in this work,
focuses instead on finding a set of low weight Pauli operators which reproduce all of the local (anti)-commutation relations of the fermionic ``edge'' ($E_{jk}$) and ``vertex'' ($V_j$) operators --- which are most concisely defined in terms of Majorana operators $\gamma_j := a_j+a_j^{\dagger}$ and $\bar{\gamma}_j := \frac{a_j-a_j^{\dagger}}{i}$:
%
%
\begin{equation}
E_{jk} := -i \gamma_j \gamma_k \;,\; V_j =: -i \gamma_j \bar{\gamma}_j.
\end{equation}
These operators are hermitian, traceless, self inverse and $E_{jk}=-E_{kj}$. The local (anti)-commutation relation which fermion to qubit mappings of this second kind aim to reproduce is that pairs of operators anti-commute if and only if they share a vertex. More precisely:
\begin{equation}\label{eq:req3}
\{E_{jk},V_j\} = 0,\;\; \{E_{ij},E_{jk}\}=0,
\end{equation}
and for all $i\neq j\neq m\neq n$:
\begin{equation}\label{eq:req2}
[V_i , V_j ]=0,\;\; [E_{ij},V_m] = 0,\;\; [E_{ij},E_{mn}]=0.
\end{equation}
All even fermionic operators (ie even products of creation and annihilation operators and sums thereof) can be expressed in terms of edge and vertex operators \cite{bravyi2002fermionic}. Furthermore, all parity preserving operators are even fermionic operators and so in accordance with parity superselection all physical fermionic observables are even fermionic operators. 

Associating Pauli operators to each edge and vertex operator such that the above conditions are satisfied almost completely defines a mapping from the even fermionic operators to qubit operators. However there exists an additional non-local relation: the product of any loop of edge operators must equal the identity. More precisely
\begin{equation}\label{eq:req4}
i^{(\vert p\vert-1)} \prod_{i=1}^{(\vert p\vert-1)} E_{p_i p_{i+1}} =1 
\end{equation}
for a cyclic sequence of sites $p = \{p_1, p_2,..\}$.

Given a mapping from edge operators to Pauli operators, the expression on the left hand side of Eq.~\ref{eq:req4} also in general maps to a Pauli operator. However by restricting the mapped fermionic states to the common $+1$ eigenspace of these Pauli operators  (ie taking them to be the stabilizers) Eq.~\ref{eq:req4} becomes satisfied in that subspace.


For all prior mappings which have employed this second strategy, the stabilizers additionally generate the total fermion parity operator $\prod_i V_i$, fixing its value to $+1$, so that the mappings only admit representations of even fermionic states. The mapping given in this work is the first employing this design strategy that can avoid this side-effect. In this case one may represent states violating parity superselection, and so the full fermionic algebra admits a representation. This additional structure is completely specified by mapping a single Majorana operator to a qubit operator, satisfying appropriate (anti)-commutation relations.

\begin{table*}
\hspace{-1cm}
    \begin{tabular}{|c| c c c c c|c c c|}
        \hline
         Mapping & \cite{bravyi2002fermionic} & \cite{verstraete2005mapping,whitfield2016local} & \cite{jiang2018majorana} & \cite{steudtner2019square}
          & \cite{setia2019superfast} &  \shortstack{ even face \\number} & \shortstack{ majority \\even faces} & \shortstack{ majority \\ odd faces} \\
         \hline
        \makecell{Qubit \\ Number} & $2L(L-1)$ & $2L^2$ & $2L(L-1)$ & $2L^2-L$ & $3L^2$ & $1.5L^2 -L$ & $1.5L^2 -L -1$ & $1.5L^2 -L+1$\\
        \hline
        \makecell{Qubit to \\ Mode Ratio} & $2-\frac{2}{L}$ & $2$ & $2-\frac{2}{L}$ & $2-\frac{1}{L} $ & $3$ & $1.5-\frac{2}{L}$ & $1.5 -\frac{2}{L} - \frac{1}{2L^2} $ & $1.5 -\frac{2}{L} + \frac{1}{2L^2} $\\
        \hline
        \makecell{Max Weight \\ Hopping} & 6 & 4 & 4 & 5 & 4  & 3 & 3 & 3 \\
        \hline
        \makecell{Max Weight \\Coulomb} & 8 & 2 & 6 & 6 & 6 & 2 & 2 & 2 \\
        \hline
        \makecell{Encoded \\Fermionic \\\ Space} & Even & Full & Even & Full& Even & Full & Even & \makecell{Full \\ Plus Qubit} \\
        \hline
        \makecell{Graph \\ Geometry} & General & General & \makecell{Square \\ Lattice} & \makecell{Square \\ Lattice} & General & \makecell{Square \\ Lattice} &\makecell{Square \\ Lattice} &\makecell{Square \\ Lattice}  \\
        \hline
    \end{tabular}

    \caption{A comparison of existing local fermion to qubit mappings on an $L\times L$ lattice of fermionic modes. The mapping presented in this work is given in the three rightmost columns. Max weight Coulomb and max weight hopping denote the maximum Pauli weights of the mapped Coulomb ($a^{\dagger}_i a_ia^{\dagger}_j a_j$) and nearest neighbour hopping ($a^{\dagger}_i a_j + a^{\dagger}_j a_i$) terms respectively. Encoded fermionic space denotes whether the full or even fermionic fock space is represented. Graph geometry denotes the hopping interaction geometry which the mapping is tailored to.}\label{table:mappings}
\end{table*}

\outline{-what have you done?}

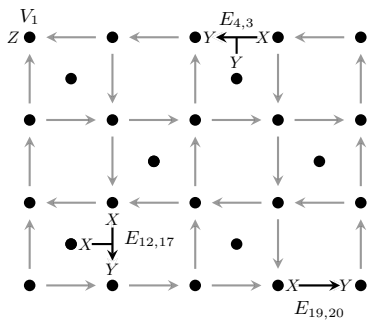
\begin{figure}[h]
\begin{center}
\begin{tikzpicture}[scale=1.1,>=stealth,thick]

    

    

\begin{scope}[shift={(5.5,0)},faint/.style={opacity=0.4}]
\foreach \x in {0,1,2,3,4}{
    \foreach \y in {0,1,2}{
        \draw[->,faint] (\x,{\y+0.5-pow(-1,\x)*0.3})--(\x,{\y+0.5+pow(-1,\x)*0.3});
        }
    }

\foreach \x in {0.5,2.5}{
    \node at (\x+1,1.5)[circle,fill=black,scale=0.5]{};
    \foreach \y in {0.5,2.5}{
        \node at (\x,\y)[circle,fill=black,scale=0.5]{};
        }
    }

\foreach \x in {0,1,2,3}{
    \foreach \y in {0,1,2,3}{
        \draw[->,faint] ({\y+0.5-pow(-1,\x)*0.3},\x)--({\y+0.5+pow(-1,\x)*0.3},\x);
        }
    }

\foreach \x in {0,1,2,3,4}{
    \foreach \y in {0,1,2,3}{
        \node at (\x,\y)[circle,fill=black,scale=0.5]{};
        }
    }
    
\node at (-.2,3)[scale=0.7]{$Z$};
\node at (0,3+0.25)[scale=0.8]{$V_1$};

\node at (0.5,0.5)[circle,fill=black,scale=0.5]{};
\node at (1.5,1.5)[circle,fill=black,scale=0.5]{};

\draw[->,white] (3-0.2,3)--(2.2,3)[];    
\node at (2.5,3.2)[scale=0.8]{$E_{4,3}$};
\node at (2.18,3)[scale=0.7]{$Y$};
\node at (3-0.18,3)[scale=0.7]{$X$};
\node at (2.5,2.7)[scale=0.7]{$Y$};
\draw[->] (3-0.25,3)--(2.25,3)[];
\draw[] (2.5,3)--(2.5,2.8)[];

\draw[<-,white] (1,0.2)--(1,1-0.2)[];
\node at (1.45,0.55)[scale=0.8]{$E_{12,17}$};
\node at (1,0.2)[scale=0.7]{$Y$};
\node at (1,1-0.2)[scale=0.7]{$X$};
\node at (0.68,0.5)[scale=0.7]{$X$};
\draw[<-] (1,0.3)--(1,1-0.3)[];
\draw[] (1,0.5)--(0.75,0.5)[];

\draw[<-,white] (4-0.2,0)--(3.2,0)[];    
\node at (3.5,-0.3)[scale=0.8]{$E_{19,20}$};
\node at (3.18,0)[scale=0.7]{$X$};
\node at (4-0.18,0)[scale=0.7]{$Y$};
\draw[<-] (4-0.25,0)--(3.25,0)[];

\node[white] at (4.2,3.2){1};
\end{scope}
\end{tikzpicture}
\end{center}
\vspace{-7mm}
\caption{ Qubit assignment, edge orientation, and examples of mapped edge and vertex operators for a $4\times 5$ square lattice. Vertices are numbered in snaking order, left to right, top to bottom.}
\label{fig:Scheme}
\end{figure}

We now proceed with a complete description of the new mapping. It suffices to define mappings from edge and vertex operators to qubit operators. Consider a square lattice of fermionic sites at the vertices. Label the faces of the lattice even and odd in a checker-board pattern. For each fermionic site $j$ associate a ``vertex'' qubit indexed by $j$.  Associate a ``face'' qubit to the odd faces. Assign an orientation to the edges of the lattice so that they circulate around the even faces clockwise or counterclockwise, alternating on every row of faces. This is illustrated in Fig.~\ref{fig:Scheme}.

Here and throughout we denote mapped operators with a tilde overscript. Let $f(i,j)$ index the unique odd face adjacent to edge $(i,j)$. For every edge $(i,j)$, with $i$ pointing to $j$, define the following mapped edge operators:
\begin{equation}
\tilde{E}_{ij} :=  \left\{ \begin{array}{rl}  X_i Y_j X_{f(i,j)} & \textrm{ $(i,j)$ oriented downwards} \\
-X_i Y_j X_{f(i,j)} & \textrm{ $(i,j)$ oriented upwards \footnote{The difference in sign introduce between the vertical up and down orientations is included to ensure that cycles around odd faces are equal to $1$ and not $-1$.}}  \\
X_i Y_j Y_{f(i,j)} & \textrm{ $(i,j)$ horizontal} \end{array} \right.
\end{equation}
\begin{equation}
    \tilde{E}_{ji} := - \tilde{E}_{ij}.
\end{equation}

For those edges on the boundary which are not adjacent to an odd face, omit the Pauli operator that would otherwise be acting on a face qubit. For every vertex $j$ define the mapped vertex operators
\begin{equation}
\tilde{V}_{j}:= Z_j
\end{equation}
 This specifies all mapped vertex and edge operators and is illustrated in Figure~\ref{fig:Scheme}.

This mapping satisfies the local (anti)-commutation relations \ref{eq:req3} and \ref{eq:req2}. The intuition is that a directed edge has an $X$ on the tail and a $Y$ on the head. Whenever the head of one edge touches the tail of another the edge operators anti-commute, while if two edges touch head to head or tail to tail they commute. By adding a qubit at some faces, and choosing an appropriate orientation for the edges, one can enforce the additional necessary anti-commutation relations at the face qubits.

Given a lattice with $M$ fermionic modes, this mapping uses fewer than $1.5M$ qubits. Furthermore this construction yields hopping and Coulomb terms with Pauli weight at most $3$. The reason that the Pauli weights and qubit numbers are so low is that the face qubits are used extremely efficiently, each one enforcing anti-commutation relations at four bounding corners.

\begin{figure}
\vspace{2mm}
    \begin{center}
    \begin{tikzpicture}[scale=1.3,>=stealth,thick]

\foreach \x in {0,1}{
    \foreach \y in {0,1}{
        \node at (\x,\y)[circle,fill=black,scale=0.5]{};
        \node at (0.5+\x-\y,-0.5+\x+\y)[circle,fill=black,scale=0.5]{};
        }
    }

\node at (0.15,0)[scale=0.8]{$Y$};
\node at (1-0.15,0)[scale=0.8]{$X$};
\node at (0.5,1.65-2)[scale=0.8]{$Y$};
\draw[->] (1-0.2,0)--(0.2,0)[];
\draw[] (.5,0)--(0.5,1.75-2)[];

\begin{scope}[rotate=180,shift={(-1,-1)}]
\node at (0.15,0)[scale=0.8]{$Y$};
\node at (1-0.15,0)[scale=0.8]{$X$};
\node at (0.5,1.65-2)[scale=0.8]{$Y$};
\draw[->] (1-0.2,0)--(0.2,0)[];
\draw[] (.5,0)--(0.5,1.75-2)[];
\end{scope}

\node at (0,0.15)[scale=0.8]{$X$};
\node at (0,1-0.15)[scale=0.8]{$Y$};
\node at (0.65-1,0.5)[scale=0.8]{$X$};
\draw[->] (0,0.25)--(0,1-0.25)[];
\draw[] (0,0.5)--(0.7-1,0.5)[];

\begin{scope}[rotate=180,shift={(-1,-1)}]
\node at (0,0.15)[scale=0.8]{$X$};
\node at (0,1-0.15)[scale=0.8]{$Y$};
\node at (0.65-1,0.5)[scale=0.8]{$X$};
\draw[->] (0,0.25)--(0,1-0.25)[];
\draw[] (0,0.5)--(0.7-1,0.5)[];
\end{scope}

\node at (2,0.5)[]{$=$};

\begin{scope}[shift={(3,0)}]
\draw (0,0)--(1,0)--(1,1)--(0,1)--cycle;

\foreach \x in {0,1}{
    \foreach \y in {0,1}{
        \node at (\x,\y)[circle,fill=black,scale=0.5]{};
        \node at (0.5+\x-\y,-0.5+\x+\y)[circle,fill=black,scale=0.5]{};
        \node at (0.15+0.7*\x,0.15+0.7*\y)[scale=0.8]{$Z$};
        
        \draw[dashed] (-0.5+1.5*\x,\y)--(1.5*\x,\y);
        \draw[dashed] (\y,-0.5+1.5*\x)--(\y,1.5*\x);
        }
    }
    
\node at (0.5,1.7-2)[scale=0.8]{$Y$};
\node at (0.5,1.5-0.2)[scale=0.8]{$Y$};
\node at (-0.3,0.5)[scale=0.8]{$X$};
\node at (1.3,0.5)[scale=0.8]{$X$};

\draw[->] (0.5,0.3) arc (270:0:0.2);

\end{scope}
\end{tikzpicture}
    \end{center}
    \vspace{-3mm}
    \caption{Non-trivial stabilizer of the encoding.}
    \label{fig:Stabilizer}
\end{figure}
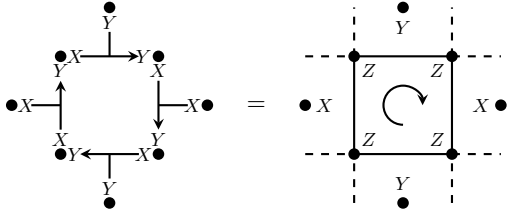

In order to satisfy Equation~\ref{eq:req4} one must project into the common $+1$ eigenspace of all loops of edge operators. 
In the case of a planar graph the loops around faces form a minimal generating set. The stabilizers associated with even faces are non-trivial and illustrated in Figure~\ref{fig:Stabilizer}, while the stabilizers associated with odd faces are equal to $1$. Therefore the number of independent constraints, dividing the Hilbert space in two, is given by the number of even faces. By a simple counting argument the dimension of the subspace to which fermionic states are mapped is thus given by:
\begin{equation}
\textrm{subspace dimension}= 2^{\textrm{M} + \textrm{OF} - \textrm{EF}}
\end{equation}
where $\textrm{M}$ is the number of fermionic modes, OF is the number of odd faces, and EF is the number of even faces.

There are three distinct cases for which the subspace dimension differs:
\begin{enumerate}[(I)]
\item There are an even number of faces, and so an equal number of even and odd faces.
\item There is one more even face than odd face.
\item There is one more odd face than even face.
\end{enumerate}
The reader may wish to examine sublattices in Figure~\ref{fig:Scheme} to develop an intuition for these cases.

In case (I) the mapping represents the full fermionic Fock space $\mathbb{F}$, with dimension $2^M$. Therefore single Majorana operators also admit a representation. It suffices to specify one Majorana, and all others may be constructed using edge and vertex operators. A Majorana operator $\gamma_{j}$ must anti-commute with all edges incident on site $j$ as well as the vertex operator $V_j$.
Any corner vertex $j$ which bounds an odd face must either have arrows pointing into it or pointing away from it.
If the arrows point into (away from) the corner, then the mapped Majorana is $\tilde{\gamma}_j=X_j \;(Y_j)$. There are two possible choices of corners. The choice is arbitrary, but once a corner is chosen then the equivalent operator at the remaining corner corresponds to a Majorana hole operator $h_i:=\gamma_i \prod_j V_j$ . 

In case (II) the dimension of the subspace is $2^{M-1}$, which is half of the full fermionic Fock space. Furthermore, up to multiplication by stabilizers, $\prod_i \tilde{V}_i =1$. Thus this mapping represents the even fermionic fock space. This is evidenced by the fact that there are no corner vertices bounding odd faces on which to define a Majorana operator.

Finally, in case (III) the dimension of the subspace is $2^{M+1}$, ie the full fermionic Fock space plus an additional ``logical'' qubit degree of freedom: $\mathbb{F} \otimes \mathbb{C}^2$. In this case there are four corner vertices on which to define a Majorana operator. These four Majoranas can be thought of as four distinct species $A_i$, $B_i$, $C_i$ and $D_i$, each introduced at a different corner and translated by edge operators to site $i$.
 Pairwise annihilation of differing species of Majorana yield three distinct vacua: $\epsilon_1$, $\epsilon_2$, $\epsilon_3$.
Identifying one species of Majorana (in this case $A$) to act as the canonical Majorana operator on the fermionic system and identity on the logical qubit, the other three species become identified with majorana hole operators multiplied by logical Pauli operators on the logical qubit.
\begin{align} 
&A_i =  \tilde{\gamma}_i  \otimes  \mathbb{I},\;\\
B_i = \tilde{h}_i  \otimes \tilde{X},\;&
C_i =  \tilde{h}_i  \otimes \tilde{Y} ,\;
D_i =  \tilde{h}_i \otimes\tilde{Z}.
\end{align}
The vacua are thus identified with applying logical Pauli operators on the logical qubit:
\begin{align} 
\epsilon_1 = \tilde{X} ,\;\; \epsilon_2 = \tilde{Y}, \;\; \epsilon_3 = \tilde{Z}. 
\end{align}

These logical Pauli operators are mapped to multi-qubit Pauli operators which span the length of the system  \footnote{\label{supMat}See Supplemental Material at [URL will be inserted by publisher] for: explicit constructions of logical qubit operators in case (III) of the mapping; an example of a variant of the mapping tailored to hexagonal lattices; and a more explicit proof of the correctness of the compact mapping.}, thus the logical qubit is topologically protected. This is highly suggestive of a connection to the toric code. 

\begin{figure}[h]
\begin{center}
\begin{tikzpicture}[scale=.9,>=stealth,thick]
\tikzmath{\w =7; \h=5; \halfh = floor(\h /2); \halfw = floor(\w/2);}
\tikzmath{ \halfwmone=\halfw-1;\halfhmone=\halfh-1; \wmone=\w-1; \hmone = \h-1;}
\foreach \x in {0,...,\halfwmone}{
        \tikzmath{\x2 = 2*\x;}
       \draw[dotted,cbPurple] (0.5+\x2+1,0)--(0.5+\x2+1,\h);
}
\foreach \y in {0,...,\halfh}{
        \tikzmath{\y2 = 2*\y;}
       \draw[dotted, cbPurple] (0,0.5+\y2)--(\w,0.5+\y2);
}
\foreach \x in {0,...,\w}{
    \foreach \y in {0,...,\h}{
        \node at (\x,\y)[circle,fill=black,scale=0.5]{};
    }
}
\begin{scope}[faint/.style={opacity=0.4}]
\foreach \x in {0,...,\wmone}{
    \foreach \y in {0,...,\halfh}{
        \tikzmath{\y2 = 2*\y;}
        \draw[->,faint] (0.2+\x,\y2)--(0.8+\x,\y2);
        \draw[<-,faint] (0.2+\x,\y2+1)--(0.8+\x,\y2+1);
    }
}
\foreach \x in {0,...,\halfw}{
    \foreach \y in {0,...,\hmone}{
        \tikzmath{\x2 = 2*\x;}
        \draw[->,faint] (\x2,0.2+\y)--(\x2,0.8+\y);
        \draw[<-,faint] (\x2+1,0.2+\y)--(\x2+1,0.8+\y);
    }
}
\end{scope}
\foreach \x in {0,...,\halfw}{
    \foreach \y in {0,...,\halfh}{
        \tikzmath{\x2 = 2*\x; \y2 = 2*\y;}
        \filldraw[cbPurple] (\x2+0.5,\y2+0.5) circle (2pt);
    }
}
\foreach \x in {0,...,\halfwmone}{
    \foreach \y in {0,...,\halfhmone}{
        \tikzmath{\x2 = 2*\x; \y2 = 2*\y;}
        \filldraw[cbPurple] (\x2+0.5+1,\y2+0.5+1) circle (2pt);
    }
}
\node at (2.2,3.2)[scale=0.8,black]{$Z$};
\node at (2.8,3.2)[scale=0.8,black]{$Z$};
\node at (2.2,3.8)[scale=0.8,black]{$Z$};
\node at (2.8,3.8)[scale=0.8,black]{$Z$};
\node at (2.2,3.2)[scale=0.8,black]{$Z$};
\node at (2.5,2.75)[scale=0.8,cbRed]{$Y$};
\node at (2.5,4.25)[scale=0.8,cbRed]{$Y$};
\node at (1.75,3.5)[scale=0.8,cbRed]{$X$};
\node at (3.25,3.5)[scale=0.8,cbRed]{$X$};

\node at (1.5,3.5)[circle,fill=cbRed,scale=0.55]{};
\node at (3.5,3.5)[circle,fill=cbRed,scale=0.55]{};
\node at (2.5,4.5)[circle,fill=cbRed,scale=0.55]{};
\node at (2.5,2.5)[circle,fill=cbRed,scale=0.55]{};

\node at (5.2,2.2)[scale=0.8,black]{$Z$};
\node at (5.8,2.2)[scale=0.8,black]{$Z$};
\node at (5.2,2.8)[scale=0.8,black]{$Z$};
\node at (5.8,2.8)[scale=0.8,black]{$Z$};
\node at (5.2,2.2)[scale=0.8,black]{$Z$};
\node at (5.35,1.75)[scale=0.8,cbBlue]{$Y$};
\node at (5.65,3.25)[scale=0.8,cbBlue]{$Y$};
\node at (4.75,2.65)[scale=0.8,cbBlue]{$X$};
\node at (6.25,2.35)[scale=0.8,cbBlue]{$X$};

\node at (4.5,2.5)[circle,fill=cbBlue,scale=0.55]{};
\node at (6.5,2.5)[circle,fill=cbBlue,scale=0.55]{};
\node at (5.5,1.5)[circle,fill=cbBlue,scale=0.55]{};
\node at (5.5,3.5)[circle,fill=cbBlue,scale=0.55]{};

\end{tikzpicture}
\end{center}
\vspace{-5mm}
\caption{The toric code ({\color{cbPurple} dotted purple}) embedded in the compact mapping. Each stabilizer is a tensor product of either a plaquette $\Pi_p$ ({\color{cbRed}red}) or star $\Pi_s$ ({\color{cbBlue}blue}) operator, with a four qubit $Z$ parity operator (black)}
\label{fig:toricCode}
\end{figure}
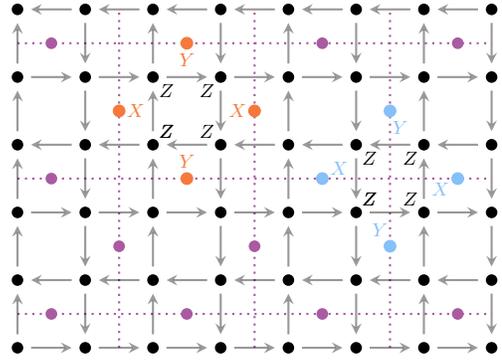

Indeed the stabilizers of the mapping presented here are tensor products of toric code star ($\Pi_S$) and plaquette  ($\Pi_P$) operators on the face qubits (up to local rotations) and  four qubit Z parity checks on the vertex qubits. This is illustrated in Figure~\ref{fig:toricCode}.

\begin{figure}[h]
\begin{center}
\begin{tikzpicture}[scale=.9,>=stealth,thick]

\tikzmath{\w =5; \h=3; \halfh = floor(\h /2); \halfw = floor(\w/2);}
\tikzmath{\halfwmone = \halfw-1; \halfhmone=\halfh-1;}
\foreach \x in {0,...,\halfwmone}{
        \tikzmath{\x2 = 2*\x;}
       \draw[dotted, cbPurple] (0.5+\x2+1,0)--(0.5+\x2+1,\h);
}
\foreach \y in {0,...,\halfh}{
        \tikzmath{\y2 = 2*\y;}
       \draw[dotted, cbPurple] (0,0.5+\y2)--(\w,0.5+\y2);
}
\foreach \x in {0,...,\w}{
    \foreach \y in {0,...,\h}{
        \node at (\x,\y)[circle,fill=black,scale=0.5]{};
    }
}

\foreach \x in {0,...,\halfw}{
    \foreach \y in {0,...,\halfh}{
        \tikzmath{\x2 = 2*\x; \y2 = 2*\y;}
        \filldraw[cbPurple] (\x2+0.5,\y2+0.5) circle (2pt);
    }
}
\foreach \x in {0,...,\halfwmone}{
    \foreach \y in {0,...,\halfhmone}{
        \tikzmath{\x2 = 2*\x; \y2 = 2*\y;}
        \filldraw[cbPurple] (\x2+0.5+1,\y2+0.5+1) circle (2pt);
    }
}

\filldraw[cbRed] (0.5,1.5) circle (4pt);
\filldraw[cbBlue] (1.5,0.5) circle (4pt);
\draw[black] (1,1) --(4,1);
\node at (1,1)[circle,fill=black,scale=0.7]{};
\node at (1,1)[circle,fill=white,scale=0.5]{};
\node at (4,1)[circle,fill=black,scale=0.7]{};
\node at (4,1)[circle,fill=white,scale=0.5]{};
\filldraw[cbRed] (4.5,1.5) circle (4pt);
\filldraw[cbBlue] (3.5,0.5) circle (4pt);

\draw[dashed,cbRed,very thick](0.5,1.5)--(4.5,1.5);
\draw[dashed,cbBlue,very thick](1.5,0.5)--(3.5,0.5);

\node at (1,1.2)[scale=0.8,black]{$Y$};
\node at (1.8,1.75)[scale=0.8,black]{$Y$};
\node at (2,1.25)[scale=0.8,black]{$Z$};
\node at (2.5,0.8)[scale=0.8,black]{$Y$};
\node at (3,1.25)[scale=0.8,black]{$Z$};
\node at (3.8,1.75)[scale=0.8,black]{$Y$};
\node at (4,1.25)[scale=0.8,black]{$X$};

\end{tikzpicture}
\end{center}
\vspace{-6mm}
\caption{A pair of Majorana particles, generated by a string of edge operators (black) in the fermionic mapping correspond to localized pairs of $e$ ({\color{cbBlue} blue}) and $m$ ({\color{cbRed} red}) particles in the toric code.}
\label{fig:toricErrors}
\end{figure}
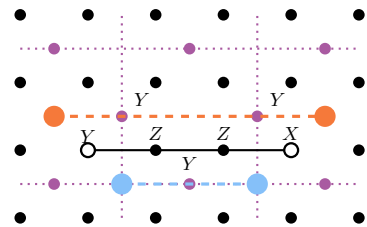

The toric code Hamiltonian:
\begin{equation}
H_{\text{toric}} = - \sum \Pi_S - \sum \Pi_P
\end{equation}
has localised electric ($e$) and magnetic ($m$) excitations, corresponding to energy contributions from $\Pi_S$ and $\Pi_P$ terms. These excitations exhibit fermionic mutual statistics and bosonic self statistics. Consequently a composite of $e$ and $m$ excitations exhibits fermionic self statistics.
The addition of the four qubit $Z$ parity operators yields a modified Hamiltonian:
\begin{equation}
H_{\text{map}} = -\sum \Pi_S \otimes Z_s -\sum \Pi_p \otimes Z_p
\end{equation}
for which specific localized pairings of $e$ and $m$ particles no longer cause an energy penalty. More concretely, any path of edge operators $E_{ij} = -i \gamma_i \gamma_j$ corresponds to the creation of a pair of Majoranas at either end of the path. Each of these Majoranas give rise to bound pairs of $e$ and $m$ particles in $H_{\text{toric}}$ which have no energy penalty in $H_{\text{map}}$ and exhibit fermionic exchange statistics. This is illustrated in Figure~\ref{fig:toricErrors}. 

 In this sense the mapping presented here leverages the  topological order of the toric code to generate non-local exchange statistics. Similar connections appear in other fermionic mappings \cite{bravyi2002fermionic, verstraete2005mapping, jiang2018majorana}, but not as explicitly.

\outline{-that was a lot to take in, can you recap?}

The fermion to qubit mapping presented here constitutes a significant improvement on both the mode to qubit ratio and the Pauli weights of local operators. For near term quantum computing hardware, such gains are essential. See Ref.~\cite{epsilonCircuts} for an example how this specific mapping can yield improvements on quantum simulation techniques.  Additionally, this mapping has error mitigating properties which we discuss in concurrent work \cite{Error-Mapping}. The design principles outlined for this encoding can also be applied to other lattice types. In the supplementary material we include a similar mapping for a hexagonal lattice. 

 We would like to thank Johannes Bausch, Toby Cubitt, Laura Clinton, Raul Santos and Tom Scruby for their helpful discussions and proofreading.

\nocite{bravyi2002fermionic}
\nocite{Ottesen1995}

\bibliography{EncodingsBib}

\end{document}


\title{A Compact Fermion To Qubit Mapping: Supplementary Material}
\maketitle

\section{Details of Logical Qubit Operators in Case III}
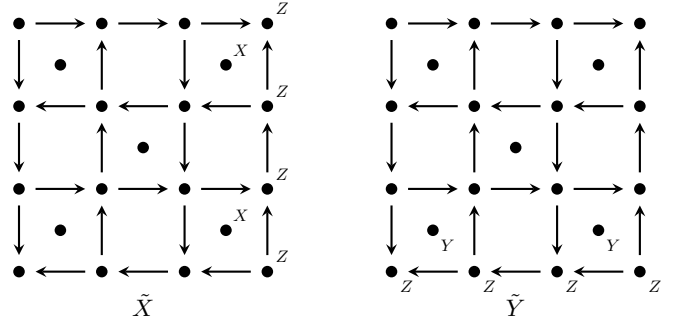
\begin{figure}[h]
\begin{center}
\begin{tikzpicture}[scale=1.1,>=stealth,thick]
\foreach \x in {0,1,2,3}{
    \node at (3.18,0.18+\x)[scale=0.7]{$Z$};
    \foreach \y in {0,1,2,3}{
        \node at (\x,\y)[circle,fill=black,scale=0.5]{};
        }
    }
    
\node at (2.5+.18,.5+.18)[scale=0.7]{$X$};
\node at (2.5+.18,2.5+.18)[scale=0.7]{$X$};

\foreach \x in {0,1,2}{
    \foreach \y in {0,2}{
        \draw[<-] (0.2+\x,\y)--(0.8+\x,\y);
        \draw[<-] (\y,0.2+\x)--(\y,0.8+\x);
        \draw[->] (0.2+\x,1+\y)--(0.8+\x,1+\y);
        \draw[->] (1+\y,0.2+\x)--(1+\y,0.8+\x);
        }
    }
    
\node at (2.5,0.5)[circle,fill=black,scale=0.5]{};
\node at (2.5,2.5)[circle,fill=black,scale=0.5]{};
\node at (0.5,0.5)[circle,fill=black,scale=0.5]{};
\node at (0.5,2.5)[circle,fill=black,scale=0.5]{};
\node at (1.5,1.5)[circle,fill=black,scale=0.5]{};

\node at (1.5,-0.4)[]{$\tilde{X}$};
    
\begin{scope}[shift={(4.5,0)}]
\foreach \x in {0,1,2,3}{
    \node at (.18+\x,-0.18)[scale=0.7]{$Z$};
    \foreach \y in {0,1,2,3}{
        \node at (\x,\y)[circle,fill=black,scale=0.5]{};
        }
    }
    
\node at (0.5+.18,0.5-.18)[scale=0.7]{$Y$};
\node at (2.5+.18,0.5-.18)[scale=0.7]{$Y$};
    
\foreach \x in {0,1,2}{
    \foreach \y in {0,2}{
        \draw[<-] (0.2+\x,\y)--(0.8+\x,\y);
        \draw[<-] (\y,0.2+\x)--(\y,0.8+\x);
        \draw[->] (0.2+\x,1+\y)--(0.8+\x,1+\y);
        \draw[->] (1+\y,0.2+\x)--(1+\y,0.8+\x);
        }
    }
    
\node at (2.5,0.5)[circle,fill=black,scale=0.5]{};
\node at (2.5,2.5)[circle,fill=black,scale=0.5]{};
\node at (0.5,0.5)[circle,fill=black,scale=0.5]{};
\node at (0.5,2.5)[circle,fill=black,scale=0.5]{};
\node at (1.5,1.5)[circle,fill=black,scale=0.5]{};

\node at (1.5,-0.4)[]{$\tilde{Y}$};
\end{scope}
\end{tikzpicture}
\end{center}
\caption{The logical $X$ and $Y$ operators on the extra logical qubit in case (III)}
\label{fig:XandY}
\end{figure}

As discussed in the main text, in case (III) the encoded subspace consists of the full fermionic Fock space plus an additional logical qubit degree of freedom ($\mathbb{F} \otimes \mathbb{C}^2$) and the square lattice has four corners from each of which either all incident edges point into or away. Applying an $X$ Pauli operator at a corner where edges point away (and $Y$ where edges pointing into) must correspond to applying an encoded Majorana or hole operator ($h_i := \gamma_i \prod_k V_k$) --- as these are the only fermionic operators satisfying the resulting (anti-)commutation relations at that site --- along with potentially some additional action on the logical qubit space. We define the four operators $A_i, B_i, C_i$ and $D_i$, each corresponding to a distinct corner, as the application of an $X$ (or $Y$) at that corner followed by the application of a sequence of edge operators from the corner to site $i$. These operators satisfy the following  (anti-)commutation relations $\forall i\neq j$ and $\forall M~\neq~ M'~\in~\{A,B,C,D\}$:
\begin{align}
\{M_i, M_j\} = [M_i, M'_j] =\{M_i, M'_i\}=0 
\end{align}
An assignment of Majorana, hole and Pauli operators to these $M$ which satisfy these commutation relations is:
\begin{align} 
&A_i =  \tilde{\gamma}_i  \otimes  \mathbb{I},\;\\
B_i = \tilde{h}_i  \otimes \tilde{X},\;&
C_i =  \tilde{h}_i  \otimes \tilde{Y} ,\;
D_i =  \tilde{h}_i \otimes\tilde{Z}.
\end{align}
One can retrieve the isolated logical Pauli operators by taking the product of select pairs:
\begin{align}
C_i D_i &= i \mathbb{I} \otimes \tilde{X}\\
D_i B_i &= i \mathbb{I} \otimes \tilde{Y}\\
B_i C_i &= i \mathbb{I} \otimes \tilde{Z}
\end{align}
Such products correspond to string-like operations extending the length of the lattice, from one corner to another. Examples of these operators are illustrated in Figure~\ref{fig:XandY}. Note that if one treats one of these operators as a stabilizer, then one restricts to the full fermionic code space without an extra logical qubit. In this case, as one should expect, there are only two corners in which to inject a Majorana, since injecting a Majorana at either of the other two corners would anti-commute with the chosen stabilizer. These two corners correspond to the Majorana and its hole counterpart, as in case (I) where there are only two odd corners.

\section{Hexagonal Lattice Mapping}
Hexagonal lattices admit a similar construction to that described in the main text for square lattices. Again we follow the scheme of orienting the edges, introducing ansatz edge operators with an $X$ at the tail and a $Y$ at the head, and then introducing interactions on face qubits to satisfy the anti-commutation relations.

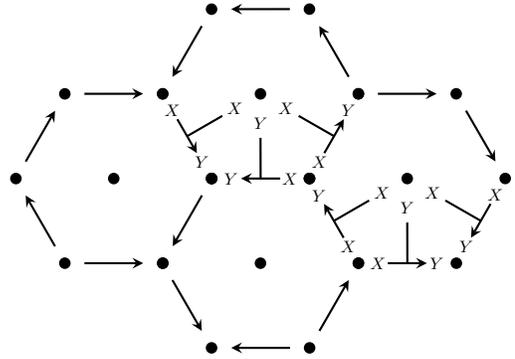
\begin{figure}[h]
\center
\begin{tikzpicture}[scale=1.3,thick,>=stealth]
\node (a) at (0,0)[circle,fill=black,scale=0.5]{};
\node (cen) at (60:1)[circle,fill=black,scale=0.5]{};
\node (b) at (120:1)[circle,fill=black,scale=0.5]{};
\node (c) at ($(b)!1!120:(a)$)[circle,fill=black,scale=0.5]{};
\node (d) at ($(c)!1!120:(b)$)[circle,fill=black,scale=0.5]{};
\node (e) at ($(d)!1!120:(c)$)[circle,fill=black,scale=0.5]{};
\node (f) at ($(e)!1!120:(d)$)[circle,fill=black,scale=0.5]{};

\draw[<-] ($(a)!0.2!(f)$)--($(a)!0.8!(f)$);
\draw[<-] ($(a)!0.2!(b)$)--($(a)!0.8!(b)$);
\draw[->] ($(f)!0.2!(e)$)--($(f)!0.8!(e)$);

\begin{scope}[shift={($(0,0)!1!120:(1,0)-(1,0)$)}]
\node (a) at (0,0)[circle,fill=black,scale=0.5]{};
\node (cen) at (60:1)[circle,fill=black,scale=0.5]{};
\node (b) at (120:1)[circle,fill=black,scale=0.5]{};
\node (c) at ($(b)!1!120:(a)$)[circle,fill=black,scale=0.5]{};
\node (d) at ($(c)!1!120:(b)$)[circle,fill=black,scale=0.5]{};
\node (e) at ($(d)!1!120:(c)$)[circle,fill=black,scale=0.5]{};
\node (f) at ($(e)!1!120:(d)$)[circle,fill=black,scale=0.5]{};

\draw[->] ($(a)!0.2!(f)$)--($(a)!0.8!(f)$);
\draw[->] ($(a)!0.2!(b)$)--($(a)!0.8!(b)$);
\draw[<-] ($(f)!0.2!(e)$)--($(f)!0.8!(e)$);
\draw[->] ($(b)!0.2!(c)$)--($(b)!0.8!(c)$);
\draw[->] ($(c)!0.2!(d)$)--($(c)!0.8!(d)$);

\end{scope}

\begin{scope}[shift={($(0,0)!1!60:(1,0)+(1,0)$)}]
\node (a) at (0,0)[circle,fill=black,scale=0.5]{};
\node (cen) at (60:1)[circle,fill=black,scale=0.5]{};
\node (b) at (120:1)[circle,fill=black,scale=0.5]{};
\node (c) at ($(b)!1!120:(a)$)[circle,fill=black,scale=0.5]{};
\node (d) at ($(c)!1!120:(b)$)[circle,fill=black,scale=0.5]{};
\node (e) at ($(d)!1!120:(c)$)[circle,fill=black,scale=0.5]{};
\node (f) at ($(e)!1!120:(d)$)[circle,fill=black,scale=0.5]{};

\draw[<-] ($(e)!0.2!(d)$)--($(e)!0.8!(d)$);
\draw[<-] ($(d)!0.2!(c)$)--($(d)!0.8!(c)$);

\node at ($(a)!0.2!(b)$)[scale=0.7]{$X$};
\node at ($(b)!0.2!(a)$)[scale=0.7]{$Y$};
\draw[->] ($(a)!0.3!(b)$)--($(a)!0.7!(b)$);
\node at ($(cen)!0.3!-30:(a)$)[scale=0.7]{$X$};
\draw ($(a)!0.5!(b)$)--($(cen)!0.45!-30:(a)$);

\node at ($(f)!0.2!(a)$)[scale=0.7]{$Y$};
\node at ($(a)!0.2!(f)$)[scale=0.7]{$X$};
\draw[<-] ($(f)!0.3!(a)$)--($(f)!0.7!(a)$);
\node at ($(cen)!0.3!30:(a)$)[scale=0.7]{$Y$};
\draw ($(a)!0.5!(f)$)--($(cen)!0.45!30:(a)$);
\node at ($(f)!0.2!(e)$)[scale=0.7]{$Y$};
\node at ($(e)!0.2!(f)$)[scale=0.7]{$X$};
\draw[<-] ($(f)!0.3!(e)$)--($(f)!0.7!(e)$);
\node at ($(cen)!0.3!90:(a)$)[scale=0.7]{$X$};
\draw ($(e)!0.5!(f)$)--($(cen)!0.45!90:(a)$);
\end{scope}

\begin{scope}[shift={($2*cos(30)*(0,1)$)}]
\node (a) at (0,0)[circle,fill=black,scale=0.5]{};
\node (cen) at (60:1)[circle,fill=black,scale=0.5]{};
\node (b) at (120:1)[circle,fill=black,scale=0.5]{};
\node (c) at ($(b)!1!120:(a)$)[circle,fill=black,scale=0.5]{};
\node (d) at ($(c)!1!120:(b)$)[circle,fill=black,scale=0.5]{};
\node (e) at ($(d)!1!120:(c)$)[circle,fill=black,scale=0.5]{};
\node (f) at ($(e)!1!120:(d)$)[circle,fill=black,scale=0.5]{};

\draw[->] ($(e)!0.2!(d)$)--($(e)!0.8!(d)$);
\draw[->] ($(d)!0.2!(c)$)--($(d)!0.8!(c)$);
\draw[->] ($(c)!0.2!(b)$)--($(c)!0.8!(b)$);

\node at ($(a)!0.2!(b)$)[scale=0.7]{$Y$};
\node at ($(b)!0.2!(a)$)[scale=0.7]{$X$};
\draw[<-] ($(a)!0.3!(b)$)--($(a)!0.7!(b)$);
\node at ($(cen)!0.3!-30:(a)$)[scale=0.7]{$X$};
\draw ($(a)!0.5!(b)$)--($(cen)!0.45!-30:(a)$);

\node at ($(f)!0.2!(a)$)[scale=0.7]{$X$};
\node at ($(a)!0.2!(f)$)[scale=0.7]{$Y$};
\draw[->] ($(f)!0.3!(a)$)--($(f)!0.7!(a)$);
\node at ($(cen)!0.3!30:(a)$)[scale=0.7]{$Y$};
\draw ($(a)!0.5!(f)$)--($(cen)!0.45!30:(a)$);

\node at ($(f)!0.2!(e)$)[scale=0.7]{$X$};
\node at ($(e)!0.2!(f)$)[scale=0.7]{$Y$};
\draw[->] ($(f)!0.3!(e)$)--($(f)!0.7!(e)$);
\node at ($(cen)!0.3!90:(a)$)[scale=0.7]{$X$};
\draw ($(e)!0.5!(f)$)--($(cen)!0.45!90:(a)$);
\end{scope}
\end{tikzpicture}
\caption{Edge orientation, qubit placement and edge operators for the hexagonal lattice mapping.} \label{fig:HexScheme}\label{fig:HexEncoding}
\end{figure}

In the hexagonal lattice every edge, except for the bottom edge of every face, is oriented clockwise on even columns of faces and counterclockwise on odd columns, as illustrated in Figure~\ref{fig:HexScheme}. This ensures that heads touch tails for all edges except for the bottom edge of every hexagon. A qubit is introduced at every face.

For every bottom edge of every face $f$, the edge operator acts on the face qubit of $f$ with $Y_f$. The two edges adjacent to this bottom edge and also adjacent to $f$ act on the face qubit of $f$ with $X_f$.  In this way all anti-commutation relations are satisfied. Just as in the square lattice case, the vertex operators are $Z$ operators on the vertex qubit. We illustrate this construction in Figure~\ref{fig:HexEncoding}.

Once again, the stabilizers of this mapping are cycles. However in this case there are no trivial cycles, so there is a stabilizer generator for every face. This implies that the code space is the full fermionic space. One again, single fermion operators may be injected into the code at those vertices from which edges are either uniformly pointing towards or away. This mapping yields hopping and coulomb terms with Pauli weight at most 3. With $M$ modes, this mapping uses fewer than $1.5M$ qubits.

Similar generalizations may be applied to other lattices.

\section{Proof of Correctness of Square Lattice Fermionic Mappings}
For the sake of completeness we include here a more explicit proof of our claims. The arguments presented here should be relatively familiar to those readers acquainted with fermionic mappings.
\begin{definition}
A \textit{mapping} from some Hilbert space $H_1$
to another Hilbert space $H_2$ is an isometry \footnote{$J^{\dagger}J = \mathbb{I}_{H_1}$ and $J J^{\dagger}= \textrm{Proj}( \textrm{Image}(J)) $}  $J: H_1 \rightarrow H_2$, as defined in \cite[Sec.~5]{bravyi2002fermionic}. 
\end{definition} 
Let $f_m = \mathbb{C}^m$ correspond to an $m$ mode single fermion Hilbert space. The $n$ fermion Fock space on $m$ modes is defined as $$\land^n f_m = \textrm{span}\left(\frac{1}{n!}\sum_{\pi \in S_n} \sgn(\pi) \bigotimes_{i=1}^n \psi_{\pi(i)} : \psi_j\in f_m \right),$$ and the full fermionic Fock space of $m$ modes is $F_m = \bigoplus_{n=0}^m (\land^n f_m).$ The dimension of $F_m$ is $2^m$. Let $F_m^E = \bigoplus_{n \in \textrm{even}}^m (\land^n f_m)$ and $F_m^O = \bigoplus_{n \in \textrm{odd}}^m (\land^n f_m)$. If $m$ is even the dimension of $F_m^E$  is $2^{m-1}$.

In the following we denote with $L(S)$ to be the Linear operators on a space $S$, which form a $C^\ast$-algebra in the case where $S$ is a complex Hilbert space. 

\begin{definition}[Fermionic Mapping]\label{def:embedding}
    A fermionic mapping is an isometry $J: F_m \longrightarrow H$, where $H$ is a qubit Hilbert space, i.e.\ $J$ satisfies the property $J^\dagger J = \mathbb{I}_{F_m}$
\end{definition}

\begin{proposition}[\cite{bravyi2002fermionic}]\label{prop:starHom}
A $\ast$-isomorphism $\mu: L(H_1) \rightarrow L(H_2)$ induces an mapping $J:H_1 \longrightarrow H_2$ which is unique up to a global phase.
An operator $X\in L(H_1)$ is represented via the map $\mu(X) = J X J^{\dagger}$. 
\end{proposition}

\begin{proposition}[\cite{bravyi2002fermionic}]\label{prop:gens}
The algebra $L(F_m^E) \oplus L(F_m^O)$ is generated by the edge and vertex operators $E_{ij}$ and $V_i$.
The algebra $L(F_m)$ is generated by a single Majorana operator $\gamma$ and the edge and vertex operators $E_{ik}$ and $V_i$.
\end{proposition}

\begin{theorem}
A square $(L_1 \times L_2)$ lattice mapping with an even number of faces $n_F = (L_1-1)(L_2-1)$ (case I) is a fermionic mapping $J: F_m \rightarrow \mathcal{L}$, where $m= L_1L_2$ and 
$\mathcal{L} \subset (\mathbb{C}^2)^{m+n_F/2}$.
\end{theorem}
\begin{proof}
The mapping employs $m + n_F/2$ qubits. The number of non-trivial stabilizers is $n_F/2$. Let $\mathcal{L} \subset H=(\mathbb{C}^2)^{m+n_F/2}$ be the joint $+1$ eigenspace of these non-trivial stabilizers, then $\dim(\mathcal{L})=2^m = \dim(F_m)$.

If edge operators $E_{jk}$ and vertex operators $V_k$ satisfy the anti-commutation relations 
\begin{equation}\label{eq:req3}
\{E_{jk},V_j\} = 0,\;\; \{E_{ij},E_{jk}\}=0,
\end{equation}
and for all $i\neq j\neq m\neq n$:
\begin{equation}\label{eq:req2}
[V_i , V_j ]=0,\;\; [E_{ij},V_m] = 0,\;\; [E_{ij},E_{mn}]=0.
\end{equation}
and loop condition
\begin{equation}\label{eq:req4}
i^{(\vert p\vert-1)} \prod_{i=1}^{(\vert p\vert-1)} E_{p_i p_{i+1}} =1 
\end{equation}
 then they generate the even fermionic algebra $L(F_m^E)$.
In order to generate the entire fermionic algebra $L(F_m)$, we need an additional generator; for instance a single Majorana operator $\gamma_i$, which together with a vertex operator $V_i\gamma_i \propto \bar\gamma_i$ generates the entire algebra $L(F_m)$.

By inspection the Pauli operators $\tilde{E}_{jk}$ and $\tilde{V}_k$ satisfy relations \ref{eq:req2} and \ref{eq:req3}.
Furthermore, since the number of faces is even there are always exactly two corner faces of the lattice which are odd according to the checker-board labelling.
Choose a corner vertex $c$, associated with an odd corner face, and define an encoded Majorana operator
\[
\tilde{\gamma}_c = \begin{cases}
X_c & \text{arrows pointing into corner, or} \\
Y_c & \text{otherwise.}
\end{cases}
\]
Then $\{ \tilde\gamma_c, \tilde E_{jk} \} = \{ \tilde\gamma_c, \tilde V_k \} = 0$.

By inspection we see that $\tilde E_{jk}$, $\tilde V_k$ and the additional Majorana $\tilde \gamma_c$ commute with the stabilizers of the mapping; as such, restriction to the joint $+1$ eigenspace $\mathcal L$ of the stabilizers---denoted $\cdot|_\mathcal L$---retains all their algebraic properties. Furthermore, the operators $\tilde E_{jk}|_\mathcal L$ satisfy Equation~\ref{eq:req4}.
Thus by the same argument as in \cite[Sec.~8]{bravyi2002fermionic}, the identification
\[
    E_{jk} \longmapsto \tilde E_{jk}|_\mathcal L
    \quad
    V_k \longmapsto \tilde V_j|_\mathcal L
    \quad
    \gamma_c \longmapsto  \tilde\gamma_c|_\mathcal L
\]
extends to a $\ast$-homomorphism $\mu : L(F_m) \longrightarrow L(\mathcal L)$. 
Since the CAR algebra is unique up to an isomorphism \cite{Ottesen1995}, and $\dim(\mathcal{L}) = \dim(F_m)$, the mapping is an isomorphism. The claim follows by \ref{prop:starHom}.
\end{proof}

\begin{theorem}
A square $(L_1 \times L_2)$ lattice mapping with an odd number of faces $n_F=(L_1-1)(L_2-1)$, and an extra even face (case II)  describes an mapping $J: F_m^E \rightarrow \mathcal{L}$, where $m= L_1L_2$ and 
$\mathcal{L} \subset (\mathbb{C}^2)^{m+ \lfloor n_F/2 \rfloor}$.
\end{theorem}

\begin{proof}
The mapping employs $m + \lfloor n_F/2 \rfloor$ qubits. The number of non-trivial stabilizers is $\lceil n_F/2 \rceil $. Let $\mathcal{L} \subset (\mathbb{C}^2)^{m+\lfloor n_F/2 \rfloor}$ be the $+1$ eigenspace of the non-trivial stabilizers, then $\dim(\mathcal{L})=2^{m +  \lfloor n_F/2 \rfloor - \lceil n_F/2 \rceil }= 2^{(m-1)}$. Since $n_F$ is odd, it must be the case that $L_1-1$ and $L_2-1$ are odd, and so $m$ is even. Thus $\dim(F_m^E) = 2^{m-1} = \dim(\mathcal{L})$.

$L(F_m^E) \oplus L(F_m^O)$ is generated by $E_{ij}$ and $V_i$. However under the algebraic constraint that $\prod_i V_i =\mathbb{I}$, the algebra only generates $L(F_m^E)$. Thus we have a mapping $\mu: L(F_m^E) \rightarrow L(\mathcal{L})$ by $\mu(E_{ij})= \tilde{E}_{ij} \vert_{\mathcal{L}}$ and $\mu(V_i)=\tilde{V}_i \vert_{\mathcal{L}}$. By a similar argument to the previous theorem, $\mu$ is a *-isomorphism, and therefore specifies an mapping $J: F_m^E \rightarrow \mathcal{L}$.
\end{proof}

\begin{theorem}
A square $(L_1 \times L_2)$ lattice mapping with an odd number of faces $n_F=(L_1-1)(L_2-1)$, and an extra odd face (case III), is a fermionic mapping $J: \mathbb{C}^2 \otimes F_m \rightarrow \mathcal{L}$, where $m= L_1L_2$ and 
$\mathcal{L} \subset (\mathbb{C}^2)^{m+ \lceil n_F/2 \rceil}$.
\end{theorem}

\begin{proof}
The mapping employs $m + \lceil n_F/2 \rceil$ qubits. The number of non-trivial stabilizers is $\lfloor n_F/2 \rfloor $. Let $\mathcal{L} \subset (\mathbb{C}^2)^{m+\lceil n_F/2 \rceil}$ be the $+1$ eigenspace of the non-trivial stabilizers, then $\dim(\mathcal{L})= 2^{(m+1)}= \dim(\mathbb{C}^2 \otimes F_m^E)$.

There are four corner faces of the lattice which are odd according to the checker-board labelling. One may choose a corner vertex $c$ and define an encoded operator acting trivially on the logical qubit, and as a Majorana on the logical fermionic space: $\tilde{\mathbb{I}}\otimes \tilde{\gamma}_c = X_c \textrm{ or } Y_c$ depending on if the arrows point into or respectively away from that corner. 
The other three corners, which we label $c_x,c_y,c_z$, may then be defined as logical Pauli operators combined with logical hole operators:
\begin{align*}
    \tilde{X}\otimes \tilde{h}_{c_x} &= X_{c_x} \textrm{ or } Y_{c_x} \\
    \tilde{Y}\otimes \tilde{h}_{c_y} &= X_{c_y} \textrm{ or } Y_{c_y} \\
    \tilde{Z}\otimes \tilde{h}_{c_z} &= X_{c_z} \textrm{ or } Y_{c_z}
\end{align*}
here the choice of which corners to associate with which logical Paulis is a matter of convention, and the choice of $X$ or $Y$ will again depend on if the arrows point into or respectively away from that corner. 

By Proposition~\ref{prop:gens}, the operators $\mathbb{I} \otimes \gamma_j$, $X \otimes h_j$ and $Y \otimes h_j$ , along with the edge and vertex operators, generate $L(\mathbb{C}^2 \otimes F_m)$. Define the mapping $\mu: L(\mathbb{C}^2 \otimes F_m) \rightarrow L(\mathcal{L})$ by $\mu(\mathbb{I} \otimes \gamma_c) = \tilde{I} \otimes \tilde{\gamma_c} \vert_{\mathcal{L}}$ etc. By a similar argument to the previous theorems, $\mu$ is a *-isomorphism, and therefore specifies an mapping $J: \mathbb{C}^2 \otimes F_m \rightarrow \mathcal{L}$.

\end{proof}

\bibliography{EncodingsBib}